\documentclass[]{aa}

\usepackage{txfonts,graphicx,rotating,natbib}
\bibpunct{(}{)}{;}{a}{}{,}

\newcommand{\oder}[2]{{{\rm d}\,#1\over{\rm d}\,#2}}
\newcommand{\pd}[2]{\,{\partial\,#1\over\partial\,#2}\,}

\begin{document}

\title{Laboratory measurement of optical constants of solid SiO and application to
circumstellar dust}


\author{Steffen Wetzel\inst{1} \and Markus Klevenz\inst{1,2} \and Hans-Peter Gail\inst{2}
\and Annemarie Pucci\inst{1} \and Mario Trieloff\inst{3}
}

\institute{
Universit\"at Heidelberg, Kirchhoff-Institut f\"ur Physik, Im Neuenheimer Feld 227,
           69120 Heidelberg, Germany 
\and
Universit\"at Heidelberg, Zentrum f\"ur Astronomie, Institut f\"ur Theoretische
           Astrophysik, Albert-Ueberle-Str. 2,
           69120 Heidelberg, Germany 
\and
Universit\"at Heidelberg, Institut f\"ur Geowissenschaften, Im Neuenheimer
           Feld 234--236, 69120 Heidelberg, Germany
  }

\offprints{\tt gail@uni-heidelberg.de}

\date{Received date ; accepted date}

\abstract
{Silicate minerals belong to the most abundant solids that form in 
cosmic environments. Their formation requires that a sufficient 
number of oxygen atoms per silicon atom is freely available. For the standard
cosmic element mixture this can usually be taken for granted, but this becomes
a problem at the transition from the oxygen rich chemistry of M-stars to the
carbon rich chemistry of  C-stars. In the intermediate type S-stars most of the
oxygen and carbon is consumed by formation of CO and SiO molecules, and 
left-over oxygen to build SiO$_4$-tetrahedrons in solids becomes scarce. 
Under such conditions SiO molecules from the gas phase may condense into
solid SiO. The infrared absorption spectrum of solid SiO differs from that of
normal silicates by the absence of Si-O-Si bending modes around $18\,\mu$m
while the absorption band due to Si-O bond stretching modes at about 
$10\,\mu$m is present. Recently it has been reported (Hony et al. 2009)
that exactly this particular characteristics is observed in a number of S-star
spectra.}
{We demonstrate that this observation may be explained by the formation of
solid SiO as a major dust component at C/O abundance ratios close
to unity.
}
{The infrared absorption properties of solid SiO are determined by 
laboratory transmission measurements of thin films of SiO
produced by vapour deposition on a Si(111) wafer. From the measured spectra the 
dielectric function of SiO is derived by using a Brendel-oscillator model,
particularly suited for the representation of optical properties of amorphous 
materials. The results are used in model calculations
of radiative transfer in circumstellar dust shells with solid SiO dust in order
to determine the spectral features due to SiO dust. 
}
{Comparison of synthetic and observed spectra shows that reasonable 
agreement is obtained between the main spectral characteristics of emission
bands due to solid silicon monoxide and an emission band centred on 10\,$\mu$m, 
but without accompanying 18$\mu$m band, observed in some S-stars. We propose
that solid SiO is the carrier material of this $10\,\mu$m spectral feature.  
}
{}


\maketitle

\titlerunning{SiO condensation in space}


\section{%
Introduction}

Silicon monoxide, SiO, is one of the most abundant and most stable molecules
encountered in space. It is found to be present in many astronomical objects 
by observing its infrared molecular bands. Usually (but not exclusively) it is
found in objects where rather warm ($T>1000$\,K) gaseous material exists, for 
instance in circumstellar dust shells. At lower temperatures the SiO molecules
tend to associate with Mg, Fe, and O (available as H$_2$O) from the gas-phase
to form solid magnesium-iron-silicates -- the main components of cosmic
dust -- and some additional minor condensed phases bearing also Al and/or Ca.
Alternatively the SiO molecules may also condense to a solid material for
their own, the silicon monoxide solid. Solid SiO is a well known material of
significant technical importance, as it is widely used for optical purposes
as anti-reflection coating. Despite of its technical use, the properties of
SiO are not well known, and its lattice structure remains enigmatic.
  
In astrophysics, solid SiO has not yet been detected as a separate dust
component. In the early discussions on the nature of ISM dust it was taken into
consideration as possible ISM dust component (\citealp{Dul77}; see also 
\citealp{Mil82}, and references therein), until amorphous silicates (olivine,
pyroxene) were identified as carriers of the observed infrared features at 
9.7\,$\mu$m and $18\,\mu$m. The formation of silicon monoxide
has been thought, however, to be an important intermediate step in the 
formation of silicate dust for the following reasons:
\begin{itemize}
\item The back-bone of the silicate mineral structure are SiO$_4$-tetrahedra
with four oxygen atoms attached to a silicon atom. The silicate minerals 
detected in space are usually identified by the stretching and bending 
vibrational modes of this particular structure.
\item On the molecular level there exists no species with comparable structure
as free gas-phase species. 
\end{itemize}
It is necessary, therefore, that the initial stages of the condensation of
silicates involve some different kind of material. Because of the high
abundance of SiO molecules in matter with standard cosmic elemental abundances
it has been speculated since the early 1980s that condensation of SiO may
be the very initial step of silicate formation \citep{Nut82,Gai86,Gai98,
Gai98b,Nut06}. It is assumed that initially silicon monoxide clusters are
formed that serve as seed particles for growth of the silicates. 

There exists, however, the possibility that in a number of cases the silicon 
monoxide seeds finally grow to a dust species of their own. The formation of
silicates requires the free availability of sufficient oxygen to build the
SiO$_4$-tetrahedrons in silicates. This is usually not a problem since the
oxygen abundance in the standard cosmic element mixture is about twice the
carbon abundance and about the sixteen-fold of the silicon abundance 
\citep[cf.][]{Lod09}. Even after formation of the extremely stable CO and SiO 
molecules sufficient oxygen remains available to form the SiO$_4$-tetrahedra of
silicate minerals from SiO. However, this does not hold during the
whole lifetime of stars. During the evolution of low and intermediate mass stars
on the asymptotic giant branch (AGB) the C/O abundance ratio stepwise increases
once ``third dredge-up'' starts to operate. This drives in a number of steps the
C/O abundance ratio to values exceeding unity, thereby passing through a stage
where the C/O ratio is close to unity. During this stage the stars show the
characteristic chemical peculiarities of the S-stars. Then almost all oxygen is
used up by the formation of CO and SiO, and with increasing C/O abundance ratio
only a decreasing fraction of the SiO molecules can finally be converted into 
silicate dust because of the decreasing fraction of oxygen left over after SiO
and CO formation. Since it is possible to form a solid material with chemical
composition SiO by condensation of SiO vapour, there arises the possibility
to form such kind of material in stellar outflows from S-stars where due to
``third dredge-up'' the available  oxygen becomes scarce.

Recently some observational results for infrared emission from dust in S-stars 
have been published which show spectral characteristics that cannot be explained
by the standard amorphous silicate dust \citep{Hon09,Sac08}. For these stars the
emission feature at about 10\,$\mu$m is as strong as usual, but the concomitant 
feature of silicate dust at 18\,$\mu$m is very weak or even seems to be missing
at all. By inspection of the spectra published by \citet{Che93} one finds some 
additional objects sharing this special property. In our opinion this property
is a clear signature of the emission from solid silicon monoxide dust. In this
material one has Si-O bonds as in silicates, which give rise to the stretching 
vibrations that are the origin of the broad 10\,$\mu$m feature of amorphous
silicates, but one has no Si-O-Si bending modes as in SiO$_4$-tetrahedra which
are the origin of the 18\,$\mu$m feature of silicates. Theoretically, this
feature does not exist for solid silicon monoxide, but in practice, since solid
silicon monoxide is prone to disintegrate into silicon nano-clusters and 
SiO$_2$ \citep[e.g.][]{Hoh03,Hap04}, the 18\,$\mu$m feature is not completely
absent but weak. In S-stars with some oxygen left over after formation of CO
and SiO molecules part of the SiO may form some silicate material while the
bulk of the SiO molecules condense into solid silicon monoxide. This would give
rise to an unusually weak 18\,$\mu$m feature compared to the 10\,$\mu$m feature.

In this paper we attempt to show that the observational finding of the existence
of S-stars with strong 10\,$\mu$m emission feature and weak or absent 
18\,$\mu$m feature can be explained by condensation of solid silicon monoxide in
their outflows at C/O abundance ratios close to unity where chemically available
oxygen to form silicates is lacking. For this purpose we calculate radiative
transfer models of circumstellar dust shells with silicon monoxide dust and
compare the resulting spectra with published spectra of S-stars.

For these radiative transfer calculations one needs the dielectric function of 
solid silicon monoxide to calculate the extinction coefficient. Because this
material is used for optical purposes the complex index of refraction is known
from the middle infrared to the extreme UV and data are listed, e.g., in 
\citet{Pal85}. More recent determinations are given by \citet{Taz06}. During
the course of our laboratory studies on materials of astrophysical interest
\citep{Kle09} the far infrared optical properties have been determined with
high accuracy. We briefly describe in this paper the measurements and their
results. In particular the dielectric function is determined by fitting a
Brendel-oscillator model \citep{Brendel92} to experimental results.
This model is particularly suited to describe infrared optical properties of
amorphous materials in contrast to the widely used Lorentz-oscillator
model that is better suited for crystalline materials. The 10\,$\mu$m feature
of silicon monoxide is always diffuse like the corresponding feature of
amorphous silicate materials. The lattice structure of solid silicon monoxide is
not well known, but a crystalline phase seems not to exist.

The plan of this paper is as follows. In Sect.~\ref{SectOptPropSiO} we briefly
describe our laboratory measurements on the optical properties of solid SiO 
and the results for the dielectric function. In Section~\ref{SectResonForSiO}
we give our arguments why solid SiO could form in outflows from S-stars.
Section \ref{SectModels} describes the radiative transfer models and gives the
results.


\section{%
Optical properties of SiO}

\label{SectOptPropSiO}

To study the infrared (IR) optical properties of SiO spectroscopic transmittance
measurements were performed \textit{in situ} on condensed SiO films produced by
evaporation under UHV conditions. The dielectric function of solid SiO is
derived via Brendel-oscillator fits to the experimental IR spectra. 

\begin{figure}[t]

\includegraphics[width=1\hsize]{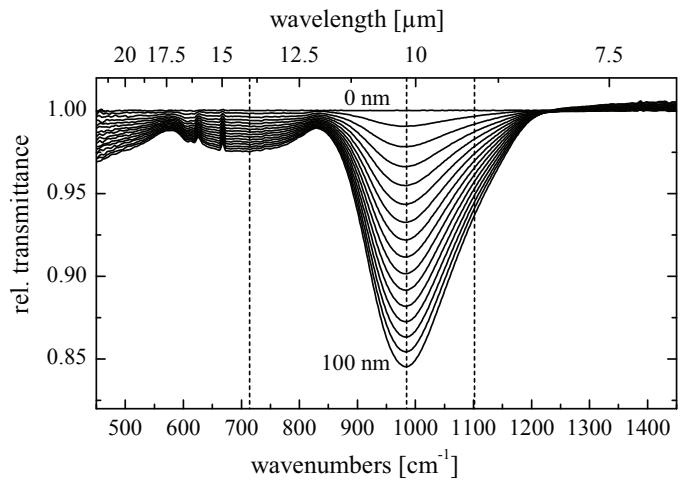}

\caption{Relative transmittance spectra of SiO on Si(111) are shown for 
increasing film thicknesses up to 100\,nm. The thickness change between two 
shown spectra is about 6.7\,nm. The strong peak at 982\,cm$^{-1}$ is attributed
to the asymmetric Si-O stretching vibration \citep{Cachard71}. For such film
thicknesses, there is clearly no shift of that peak. Narrow features in the
spectra are due to incomplete compensation of absorptions in the optical beam
path of the purged spectrometer. SiO oscillator positions are indicated by
dashed lines.}
\label{fig:thick} 
\end{figure}

\subsection{%
Experimental setup}

The experiments were performed under UHV conditions (base pressure
$<10^{-10}$\,mbar). IR spectra were taken in the range between 450\,cm$^{-1}$ 
and 5000\,cm$^{-1}$ with a Fourier-transform IR spectrometer (Bruker Tensor 27 
with mercury-cadmium-telluride detector and N$_2$ purged beam path). The IR 
spectral measurements were performed in transmittance geometry at normal 
incidence of light. A floating zone Si(111) wafer with dimensions of 
$10\times10$\,mm$^2$ and a thickness of 0.525\,mm and high resistivity was used 
as transparent substrate which was cleaned by heating up to 1000$^{\circ}$C in 
UHV to remove the natural oxide layer. Spectra were taken \textit{in situ},
i.e., during the growth of the evaporated film on the substrate (spectral 
resolution of 4\,cm$^{-1}$), and normalized to the spectrum of the bare 
Si(111) taken directly before deposition. This method allows to monitor the 
whole growth process and to detect changes due to the film growth.

Commercial SiO (Noah Technologies Corporation, Silicon Monoxide, SiO, 99.99\% 
pure, -325 mesh, CAS no. 10097-28-6) was evaporated from a tantalum Knudsen
cell heated by electron bombardment. The background pressure measured during
evaporation away from the SiO molecular beam did not exceed 
$5\times10^{-10}$\,mbar. The SiO deposition rate (about 0.8\,nm/min) was 
determined before and after the spectroscopic experiment with a quartz
crystal microbalance with a relative error up to 5\%. This error is caused by
the uncertainty in the density of the evaporated film in the thickness 
calculation for which the density of bulk SiO (2.18\,g/cm$^3$) was taken 
\citep{Hass54}. Further experimental details can be found in 
\citet{Klevenz09AS}.

\begin{figure}[t]
\includegraphics[width=1\hsize]{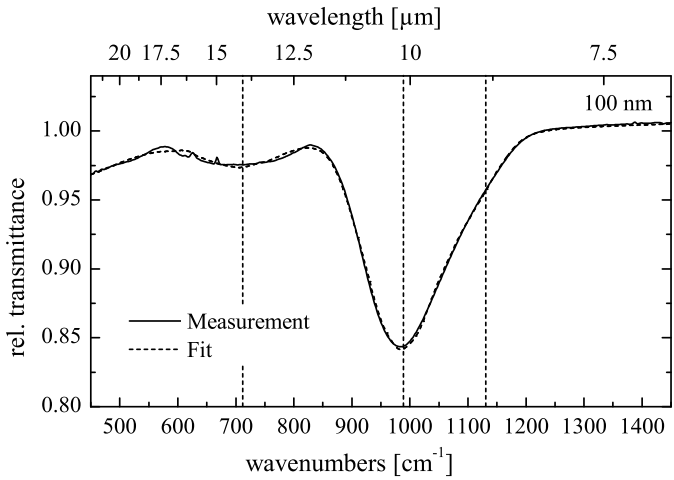}

\caption{Normal transmittance measurement of an evaporated SiO film (black solid
line) with 100\,nm thickness shown together with the best fit by a Brendel-model
of the dielectric function as described in the text. Peak positions are
indicated by dashed lines.}

\label{fig:Transmittance} 
\end{figure}

\subsection{Results}

In Fig.~\ref{fig:thick} relative transmittance spectra of SiO with increasing
film thicknesses are shown. The films were produced by SiO evaporation and
subsequent condensation on a Si(111) substrate hold at 300\,K. Two broad
features develop with growing film thickness. The stronger one at 982\,cm$^{-1}$
is assigned to the asymmetric stretching mode of the oxygen atom in a Si-O-Si 
bridge against the Si-neighbor atoms \citep{Philipp71,Chabal02,Queeney04,
Cachard71,Lehmann83,Lehmann84}. The much weaker structure at approximately 
713\,cm$^{-1}$  corresponds to the Si stretching mode, a mode with domina\-ting
Si displacement \citep{Lehmann83}, sometimes called ``bending" mode in the
literature. The third strong peak caused by the rocking mode at about 
380\,cm$^{-1}$ \citep{Philipp71}  was out of our measurement range. It is
important to note that already from 1\,nm the IR spectral features of the
evaporated SiO films do not change anymore \citep{Klevenz09PSSB}. Accordingly,
the IR dielectric function of the evaporated film already resembles that of
bulk SiO \citep{Hjortsberg80,Taz06,Klevenz09AS}.

\subsection{Dielectric function}

For thin dielectric films with the thickness $d$ of the film much smaller than
the wavelength $\lambda$ ($d\ll\lambda$), approximate formulas for the
transmittance can be derived based on the Fresnel equations \citep{Berreman63}.
For a thin layer on a thick (non-interfering) substrate the normal 
transmittance ($\phi=0^{\circ}$) is given by \citep{Lehmann88,Teschner90}
\begin{equation} 
T_{\mathrm{rel}}\approx 1-\frac{2d\frac{\omega}{c}}
{1+\sqrt{\varepsilon_\mathrm{s}}}\,\mathrm{Im}\,(\varepsilon_\mathrm{f})\,,
\end{equation}
with $\omega=2\pi c/{\lambda}$, $\varepsilon_\mathrm{s}$ as the dielectric
function of the substrate and $\varepsilon_\mathrm{f}$ as the dielectric
function of the film. In transmittance measurements under normal incidence 
($\phi=0^{\circ}$) therefore only the transverse optical (TO) modes with
frequency at the maximum of the imaginary part of the dielectric function 
$\varepsilon$ and with a dynamic dipole moment parallel to the surface can be 
observed.

To determine the dielectric function of SiO in the IR from such thin film
measurements an appropriate model for the dielectric function has to be used. A 
simple Lorentz-oscillator model can not be applied in our case due to the
amorphous character of the film. The model introduced by \citet{Brendel92}
accounts for the amorphous structure of a material by assuming that the 
different IR modes can be represented by Lorentz-oscillators, but with randomly
shifted resonance frequencies that are distributed according to a Gaussian
probability distribution. The dielectric function in the IR therefore is assumed
in this model to be given by the relation
\begin{equation}
\varepsilon(\omega)_\mathrm{IR}=\varepsilon_\infty+\sum\limits_{j=1}^N
{1\over2\pi\sigma_j}\int\limits_0^\infty{\rm d}z\,{\rm e}^{-(z-\omega_{0,j})^2/2\sigma_j^2}
{\omega_{{\rm p},j}^2\over z^2-\omega^2+{\rm i}\gamma_j\omega}\,.
\label{ModlLorentzAm}
\end{equation}
It consists of a dielectric background $\varepsilon_{\infty}$ and $N$
distributions of Lorentz oscillators with resonance frequencies $\omega_{0,j}$, 
damping constants $\gamma_{j}$, plasma frequencies $\omega_{{\rm p},j}$, and
with standard deviations $\sigma_{j}$ of the Gaussian probability distributions.

\begin{figure}[t]

\includegraphics[width=1\hsize]{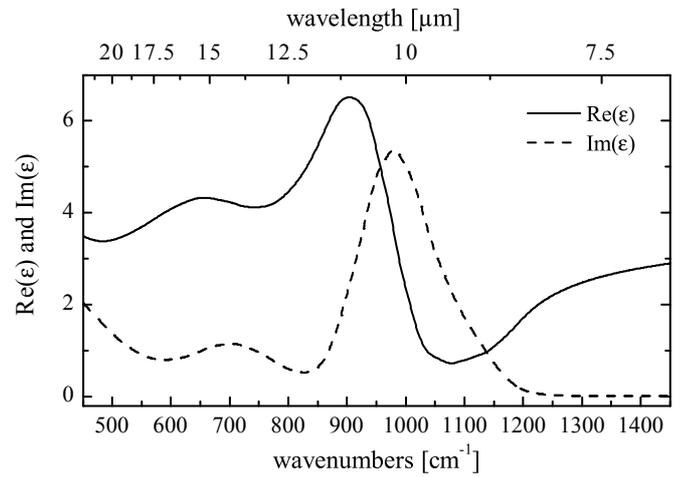}

\caption{The dielectric function obtained from the best fit with four Brendel
oscillators, as shown in Fig.~\ref{fig:Transmittance}.}
\label{fig:DielectricFunction} 
\end{figure} 

There is a close relationship of the Brendel oscillator model to the 
well known Voigt function for the profiles of velocity broadened spectral
lines. This is briefly discussed in Appendix~\ref{SectVoigt}.

The last spectrum of Fig.~\ref{fig:thick} was fitted with such a Brendel 
dielectric function using four oscillators, similar to a procedure for 
SiO$_2$\,\citep{Kirk88}, which results in very good agreement 
(see Fig.~\ref{fig:Transmittance}). We performed the spectral fits in the range 
between 500\,cm$^{-1}$ and 1500\,cm$^{-1}$ using the software package 
SCOUT  \citep{Scout}. For the high frequency limit $\varepsilon_\infty$ of the 
dielectric function a value of 3.61 was taken from literature
\citep{Hjortsberg80}. Since the oscillator damping constant $\gamma$ cannot
properly be  determined from fits to vibration spectra of disordered solids
\citep[cf.][] {Ishikawa00} we fixed the value of $\gamma$ to the resolution of
our measurement of 4\,cm$^{-1}$. Nearly the same kind of procedure is already
reported in the literature \citep{Ishikawa00,Brendel92,Grosse86,Naiman85}. The
result for the dielectric function of SiO in the IR is shown in 
Fig.~\ref{fig:DielectricFunction}, and our best fit for the oscillator
parameters is given in Table~\ref{ConstEpsSiO}.

\begin{table}

\caption{Constants of the Brendel oscillator model for the dielectric function
of amorphous materials, Eq.~(\ref{ModlLorentzAm}), for our best fit to our 
experimental results on solid SiO. All quantities are given in cm$^{-1}$ units.}

\begin{tabular}{crrrrrrrr}
\hline\hline
\noalign{\smallskip}
Osc.  &\hspace*{.6cm}& \multicolumn{1}{c}{$\omega_0$} &
\hspace*{.4cm}& \multicolumn{1}{c}{$\omega_{\rm p}$} &
\hspace*{.4cm}& \multicolumn{1}{c}{$\sigma$}&
\hspace*{.4cm}& \multicolumn{1}{c}{$\gamma$} \\
\noalign{\smallskip}
\hline 
\noalign{\smallskip}
1   &&  1101 && 305 &&  44.9 && 4 \\
2   &&  982  && 699 &&  57.4 && 4 \\
3   &&  713  && 298 &&  73.4 && 4 \\
4   &&  380  && 461 && 127.1 && 4 \\
\noalign{\smallskip}
\hline
\noalign{\smallskip}
$\varepsilon_\infty$ && 3.61 \\
\noalign{\smallskip}
\hline
\end{tabular}
\label{ConstEpsSiO}
\end{table}

These results are obtained for a substrate temperature of 300\,K. Other
substrate temperatures than 300\,K result in slightly shifted peak positions.
At a substrate temperature of 93\,K the position of the asymmetric stretching
mode is at 969\,cm$^{-1}$, while for 300\,K it is at 982\,cm$^{-1}$. For
temperatures higher than 300\,K the peak shifts to higher wavenumbers 
\citep{Kle09}. The shift to higher wavenumbers can be explained by a
decomposition into Si and SiO$_2$ that accelerates with temperature. The
effects of the substrate temperature on the condensed SiO film and the
dielectric function of SiO will be discussed in detail in a forthcoming paper.
In our present communication the dielectric function obtained at 300\,K is
used.


\section{%
Astrophysical applications}

\label{SectResonForSiO}

In this section we discuss the possibility that solid silicon monoxide might be
an important dust species in the dust shells of S-stars. 

\subsection{S-stars}

Low and intermediate mass stars from the range of initial masses below $\approx
8\,\rm M_{\sun}$ evolve till the very end of their lives to the thermally
pulsing Asymptotic Giant Branch (TP-AGB) where they are composed of an
electron-degenerated core of carbon and some oxygen, resulting from He-burning,
an overlying layer of mainly He, resulting from H-burning via the CNO-cycle, and
an enormously extended hydrogen rich envelope. On the TP-AGB the stars
alternatively burn either H in a shell source at the interface between the 
H-rich envelope and the He layer for a period of several thousand years, or He 
in a shell-source at the interface between the He layer and the carbon-oxygen 
core for a period of about 200\,yr. During the short burning phase of He (the 
thermal pulse) the He layer is for part of the time fully convective from bottom
to top and distributes freshly synthesized carbon from the He-burning shell at 
the bottom of the He layer over the whole layer. Somewhat later in the pulse
phase, the convection zone of the fully convective hydrogen envelope briefly
dips into the upper parts of the He layer and mixes some carbon-rich material
into the outer envelope. As a result, after each thermal pulse the carbon
abundance of the envelope increases stepwise. The oxygen abundance in the
envelope remains almost unchanged by this process because only very little
oxygen is synthesized by He-burning in TP-AGB stars. Thus, the carbon to oxygen
abundance ratio C/O in the envelope changes from its initial value of 
$\approx 0.5$, corresponding to the C/O ratio of the cosmic standard element
mixture \citep[see, e.g.,][]{Lod09}, to C/O ratios far exceeding a value of
unity. 

For most of the stars on the TP-AGB the C/O abundance ratio falls --- after
mixing carbon from the core into the envelope during a thermal pulse --- for one
or two inter-pulse phases into the critical range between about 0.9 and 1.0
where the chemical composition of the material in the stellar photosphere
dramatically changes from an oxygen-compound dominated composition for a C/O
abundance ratio below 0.9 and a hydrocarbon-dominated composition for a C/O
abundance ratio exceeding~1.0. This switching of the chemical composition is
brought about by the extraordinary stability of the CO molecule that almost
completely consumes by its formation the less abundant of the two elements C and
O. Only part of the TP-AGB stars leap over the critical range of C/O abundance
ratios during a single thermal pulse.

In the intermediate abundance range of C/O ratios between 0.9 and 1.0 the
stellar spectrum is dominated by  molecular bands of some low-abundance elements
that are neither seen in M-Stars (C/O ratio $\lesssim0.9$) nor in C-stars (C/O
ratio $\gtrsim1.0$). These stars are the so called S-stars. They are much less 
abundant than M-stars or C-stars since stars on the TP-AGB suffer numerous 
thermal pulses, but only during one or two inter-pulse phases they are S-stars. 

For these S-stars the material that outflows from the stellar surface neither
can form the silicate dust that is seen in circumstellar dust shells around 
M-stars, nor the graphite and SiC dust seen in dust shells around C-stars. This
is because the excess oxygen over carbon, that is not bound in CO molecules, is
too rare to combine with all of the silicon to the SiO$_4$-tetrahedrons that
form the back-bone of silicate minerals. At the same time no excess carbon is
available to form solid carbon. 

The details of the chemistry of dust formation in outflows from S-stars is 
discussed in \citet{Fer02}. Because of the very high bond energy of the SiO 
molecule --- not as high as for CO, but also exceptionally high --- the silicon
is bound in this molecule and the excess of the oxygen not bound in
SiO and CO forms H$_2$O. Some quantities of silicate dust may be formed in 
outflows from S-stars if the C/O abundance ratio is not too close to unity,
because then SiO molecules react with the left over H$_2$O and with Mg 
(and/or Fe). Indeed, weak emission bands showing the characteristic two peaks
around 9.7\,$\mu$m and 18\,$\mu$m are seen in the infrared spectrum of a number
of S-stars \citep{Che93,LLo99}. It is argued by \citet{Fer02} that also iron
dust may be formed and may be an abundant dust species for S-stars. This dust
would be hard to detect observationally for optically thin dust shells, since
its smooth and featureless extinction  would act as a more-or-less gray opacity
source. For optically thick shells it is likely that this is mistaken to be
carbon dust because of almost indistinguishable infrared extinction properties
of iron and carbon dust. 

\begin{figure}

\includegraphics[width=\hsize]{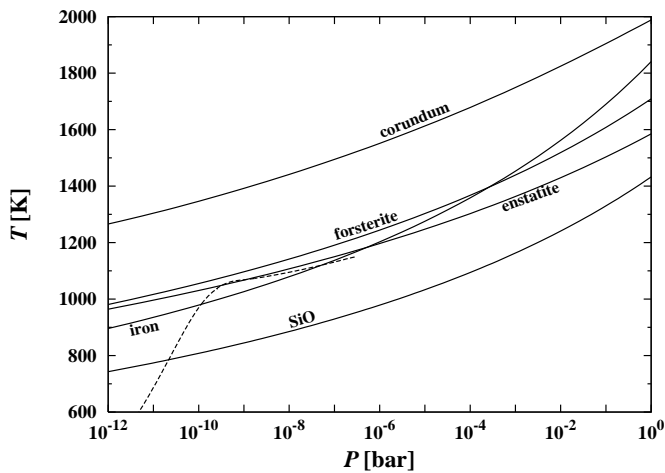}

\caption{Stability limits against vaporisation of some refractory compounds
formed from an oxygen-rich element mixture with cosmic standard element
abundances,  and of solid SiO. Dashed line: typical pressure-temperature
combinations in an outflowing gas element for a stellar wind model with
stationary outflow and $\dot M=5\times10^{-6}\, \rm M_{\sun}\,yr^{-1}$ (here
for an M-star). Most part of the dust condensation occurs around the knee where
the velocity turns from sub-sonic (horizontal part of the curve) to highly
supersonic (steep descending part of the curve) outflow. Pressures are typically
$10^{-10}$ bar in the region of main dust growth.
}

\label{FigLimCosm}
\end{figure}

\subsection{Why solid silicon monoxide?}
\label{SectSstars}

If the abundance of excess oxygen not bound in CO and SiO tends to zero for C/O 
abundance ratios approaching unity, there exists the possibility that SiO 
molecules start to condense as a solid of their own. This has not yet seriously
be taken into account as a dust component in circumstellar dust shells, though
it was discussed several times whether it could be the first condensate to form
the necessary seed particles for silicate dust formation 
\citep{Nut82,Gai86,Gai98,Gai98b,Nut06}. This is a crucial question for cosmic
dust formation since it is not possible to form the necessary seed particles
for silicate dust growth from a material with the composition and lattice
structure of one of the many silicate minerals. 

The main reason why solid SiO was not considered as one of the major dust
materials in circumstellar shells was the seemingly high vapour pressure from
which one predicts a rather low condensation temperature under the low pressure
conditions of circumstellar dust shells. New determinations of the
vapour pressure of solid SiO  \citep{Fer08,Kle09} have now shown that the older
measurement were seriously in error and strongly overestimated the vapour
pressure. The new results of \citet{Fer08} and our own results \citep{Kle09}
can be approximated by a vapour pressure formula for the equilibrium pressure 
of SiO molecules over solid silicon monoxide
\begin{equation}
p=\exp\left(-{42648\over T}+33.848\right)\,,
\end{equation}
where the pressure is in units of dyn\,cm$^{-2}$ (as usual in astrophysics). Our
results \citep{Kle09} extend the range of measured vapour pressures down to
the pressure range relevant for circumstellar dust shells. No extrapolation of
vapour pressures measured at higher temperatures in the laboratory to
significantly lower temperatures is required. The details will be discussed in 
a separate paper. 

Figure \ref{FigLimCosm} shows the resulting stability limit against evaporation
in chemical equilibrium for the astrophysically important silicates, for solid
iron, and for solid SiO, calculated by using the new results for the vapour
pressure of solid  SiO and thermochemical data from \citet{Bar95} for the other
materials, for the case of standard cosmic element abundances 
\citep{Asp05,Lod09}. At C/O abundance ratios close to unity no silicates can be
formed in S-stars because of a lack of freely available oxygen to convert a 
significant fraction of the SiO molecules from the gas phase into the 
SiO$_4$-tetrahedrons in silicate minerals (e.g., for a C/O ratio of 0.95 only
about 20\% of the Si can condense as silicate dust). Then the SiO molecules may
condense instead into solid SiO. At a typical pressure of $p=10^{-10}$\,bar in
the dust condensation layer of circumstellar dust shells 
(see Fig.~\ref{FigLimCosm}) the SiO would become stable against evaporation at
about 800\,K. This is high
enough that such dust could be observed as warm dust in a circumstellar dust
shell. Hence solid SiO is a candidate dust material to be observed in S-stars.%
\footnote{A lack of oxygen to form normal silicates may also be encountered in
massive supergiants if the outer layers are peeled off by mass-loss and
material that has burned H via the CNO-cycle appears at the surface. Then the O
abundance may drop to values of the order of the Si abundance and solid SiO 
may form also in the outflows from such stars.}

Whether this material really forms can be tested observationally. The
absorption properties  of solid SiO are sufficiently different from normal
silicate materials that they can be discriminated by their spectral features.
Amorphous olivine and pyroxene both have two strong absorption features centred
on 9.7\,$\mu$m and 18\,$\mu$m, that are usually observed as emission features
in the infrared spectra from circumstellar dust shells. For SiO the strong
feature at 18\,$\mu$m is missing because this corresponds to rocking modes in
the SiO$_4$-tetrahedron, but a strong feature at about 10\,$\mu$m is present
because this originates from Si-O bond stretching vibrations. Hence, the
presence of a strong $10\,\mu$m feature without a strong 18\,$\mu$m feature is
diagnostic for solid SiO. 

These characteristics have, indeed, been observed for a number of S-star
spectra \citep{Hon09} and one may speculate that they result from solid SiO as
a major dust component in such dust shells. In order to test this hypothesis we
have performed radiative transfer calculations using our new results for the 
dielectric function of SiO.

\begin{table}

\caption{Some data used for calculating the opacity of the dust species
considered in the model calculations and the corresponding key elements}

\begin{tabular}{lcccc}
\hline\hline
\noalign{\smallskip}
species & $A_{\mathrm{d}}$ & $\rho_{\mathrm{d}}$ &  El. & $\epsilon$ \\
\noalign{\smallskip}
\hline
\noalign{\smallskip}
SiO      &  44.09 & 2.13 & Si & $3.55\ 10^{-5}$ \\
iron     &  55.85 & 7.87 & Fe & $3.16\ 10^{-5}$ \\
olivine  & 172.2  & 3.81 & Si & $3.55\ 10^{-5}$ \\
pyroxene & 116.2  & 3.61 & Si & $3.55\ 10^{-5}$ \\
\noalign{\smallskip}
\hline
\end{tabular}
\label{TabDustSpec}
\end{table}

\subsection{Growth of silicon monoxide grains}

Now we turn to the question whether growth of silicon monoxide is kinetically
possible. Consider the simple approximation that a spherical dust grain 
co-moving with the outflowing gas collects SiO molecules from the gas phase.
For simplicity we neglect evaporation. The equation of growth of radius $a$
for a single grain is
\begin{equation}
v_\mathrm{exp}\oder ar=V_{\mathrm{d}}\,\alpha\,n_{\mathrm{gr}}\,v_{\mathrm{gr}}
\Phi(U_{\mathrm{drift}})
\label{EqGrGrSimpl}
\end{equation}
if we introduce by d$r=v_\mathrm{exp}$d$t$ the radial coordinate instead of time
as independent variable, $v_\mathrm{exp}$ being the outflow velocity. 

The meaning of the quantities in the growth equation is as follows: 
$V_{\mathrm{d}}$ is the volume of one chemical formula unit in the condensed
phase 
\begin{equation}
V_{\mathrm{d}}={A_{\mathrm{d}}m_{\mathrm{H}}\over \rho_{\mathrm{d}}}\,.
\end{equation}
Here $A_{\mathrm{d}}$ is the atomic weight of the condensed phase and 
$\rho_{\mathrm{d}}$ its mass-density. Numerical values are given in 
Table~\ref{TabDustSpec}. 

The quantity $\alpha$ is the growth coefficient. This seems to be first measured
by \citet{Gun58} who found a rather small value of $\alpha\approx4\times10^{-3}$
in the temperature range 1\,200\,K \dots\ 1\,500\,K. \citet{Roc92} made a new 
determination of the vapour pressure and of the evaporation coefficient $\alpha$
of SiO and found that their results for $\alpha$ can be approximated by 
\begin{equation}
\alpha(T)=0.1687-2.909\times10^{-4}\,T+1.373\times10^{-7}\,T^2
\end{equation} 
in the temperature range 1\,175\,K to 1\,410\,K. Essentially the same results for
$\alpha$ are found by \citet{Fer08}. Extrapolating these results down to a
temperature of 800\,K results in $\alpha\approx0.025$. This value will be used
in the following considerations.

The quantity $n_{\mathrm{gr}}$ is the particle density of the growth species.
If we  concentrate on the initial growth phase where not yet much of the growth
species is consumed from the gas phase, we can approximate $n_{\mathrm{gr}}$ by
\begin{equation}
n_{\mathrm{gr}}=\epsilon_{\mathrm{gr}}{\dot M\over1.4m_{\rm H}\,4\pi r^2 
v_\mathrm{exp}}\,,
\end{equation}
where $\epsilon_{\mathrm{gr}}$ is the element abundance of the element (Si in 
our case) that determines the gas-phase abundance of the growth species. The
second quantity on the r.h.s. is the density of H nuclides in a spherically 
symmetric, stationary outflow with velocity $v_\mathrm{exp}$ and mass-loss rate
$\dot M$. 

The quantity
\begin{equation}
v_{\mathrm{gr}}=\sqrt{kT\over 2\pi A_{\mathrm{gr}}m_{\mathrm{H}}}
\end{equation}
is the thermal velocity of the growth species (SiO molecules in our case)
and $A_{\mathrm{gr}}$ their molecular weight, while
\begin{equation}
\Phi(U)=\sqrt{1+{U^2\over16 v_{\mathrm{gr}}^2}}
\end{equation}
is a correction factor for particle drift relative to the gas with relative
velocity $U$. If $U\gg4v_{\mathrm{gr}}$ the particle drift is supersonic with
respect to the thermal velocity of the growth species (not the carrier gas!); 
then $v_{\mathrm{gr}}\,\Phi(U)\approx \frac14U$.

Typical values of all these quantities required for calculating SiO growth are
given in Table~\ref{TabDuGr}.

\begin{table}

\caption{Data used for calculation of silicon monoxide growth}

\begin{tabular}{l@{\hspace{1.cm}}l@{\hspace{1.cm}}r}
\hline
\hline
\noalign{\smallskip}
Quantity & unit & value\\
\noalign{\smallskip}
\hline
\noalign{\smallskip}
$V_{\rm d}$ & cm$^3$ & $3.46\times10^{-23}$ \\
$A_{\mathrm{gr}}$ & &  44.09\\
$\epsilon_{\mathrm{gr}}$ & & $3.55\times10^{-5}$ \\
$\epsilon_{\mathrm{d}}$ & & $5\times10^{-13}$ \\
$\alpha$ && 0.025 \\
$a_{\max}$ & $\mu$m & 0.0837 \\ 
$T_{\mathrm{c}}$ & K & 800 \\
$v_{\mathrm{th}}$ & cm\,s$^{-1}$ & $1.54\times10^4$ \\
$R_{\mathrm{c}}$ & cm & $ 3.2\times10^{13}$\\
\noalign{\smallskip}
\hline
\end{tabular}

\label{TabDuGr}
\end{table}

For the following we assume the outflow velocity $v_\mathrm{exp}$ in the region
where dust grows to be constant. This is not completely correct since dust 
condensation results in an acceleration of the wind, but the velocity increase
is moderate if we consider S-stars. It increases from a velocity of about the
sonic velocity (2\,km\,s$^{-1}$ -- 3\,km\,s$^{-1}$) at the inner edge of the 
dust shell to the final outflow velocity that is at most 10\,km\,s$^{-1}$ for
S-stars \citep[e.g.][]{Ram06}. Then integrating from the inner radius 
$R_{\rm c}$ of the dust shell, where dust commences to grow, to infinity we 
obtain
\begin{equation}
a_\infty=a_0+V_{\mathrm{d}}\,\alpha\,v_{\mathrm{gr}}\Phi\,
\epsilon_{\mathrm{gr}}\,
{\dot M\over1.4m_{\mathrm{H}}\,4\pi R_{\mathrm{c}}\,v_\mathrm{exp}^2}\,,
\end{equation}
where $a_0$ is the radius of the seed nuclei for dust growth and $a_\infty$
the grain radius at infinity. This equation requires that the grain radius  
$a_\infty$ remains smaller than the maximum radius $a_{\max}$, attained if all 
condensable material is condensed, because the consumption of the growth species 
is neglected in Eq.~(\ref{EqGrGrSimpl}). The maximum possible radius to which
a  particle may grow in the outflow is given by
\begin{equation}
{4\pi\over3}(a_{\max}^3-a_0^3)\epsilon_{\rm d}=V_{\rm d}\epsilon_{\rm gr}\,,
\label{DefMaxVolCond}
\end{equation}
if $\epsilon_{\rm d}$ is the number of dust grains per hydrogen nucleus. This
quantity is not precisely known; we assume a value of $5\times10^{-13}$ that is
a typical value found by observations \citep{Kna85}.

We can now write
\begin{equation}
a_\infty=a_0+a_{\max}\,{\dot M\over\dot M_{\rm cr}}
\end{equation}
with
\begin{equation}
\dot M_{\rm cr}={a_{\max}\,1.4m_{\mathrm{H}}\,4\pi R_{\mathrm{c}}\,v_\mathrm{exp}^2\over
V_{\mathrm{d}}\,\alpha\,v_{\mathrm{gr}}\Phi\epsilon_{\mathrm{gr}}}\,.
\end{equation}
The mass-loss rate $\dot M$ has to be smaller than $\dot M_{\rm cr}$ in order
that our assumption is valid, that there is no strong depletion of the gas phase
from condensable material.

For applying this to silicon monoxide condensation we have to specify
the values of $v_\mathrm{exp}$, $R_{\rm c}$ and $\Phi$. All other quantities
are given in Table~\ref{TabDuGr}. For the outflow velocity we assume a value
of $v_\mathrm{exp}=3\,\rm km\,s^{-1}$, slightly higher than the sound velocity
of the gas, since most of the grain growth proceeds before the gas is
accelerated to highly supersonic outflow velocities; otherwise further growth
is suppressed by rapid dilution of the gas. For $R_{\rm c}$ we assume a value
$R_{\rm c}=3\times10^{14}~\rm cm$, corresponding to about 8 stellar radii, see
Table~\ref{TabShellParm}. For the drift velocity we assume the same value as
for $v_\mathrm{exp}$ since typical drift velocities usually are of the order
of the sonic velocity of the gas, except for very high mass-loss rates where
they are much less. With these estimated values we obtain
\begin{equation}
\dot M_{\rm cr, sil}=4.5\times10^{-5}~\rm M_{\sun}\,yr^{-1}\,.
\end{equation}
This shows that significant condensation of solid silicon monoxide is 
kinetically possible, at least for mass-loss rates exceeding about $10^{-6}\,\rm 
M_{\sun}$.

The result for $\dot M_{\rm cr, sil}$ depends on the rather uncertain value of
$\alpha$ that is determined from extrapolating laboratory measured values
(which decrease with decreasing temperature) to much lower temperature. The 
value of 0.025 at 800\,K determined this way may be too low. In this case  
$\dot M_{\rm cr, sil}$ would be overestimated and growth of solid silicon
monoxide may be even more favourable as in the estimate above.


\section{%
Radiative transfer model for dust shells}
\label{SectModels}

\subsection{Opacity}
\label{SectOpac}

The spectral energy distribution of the emission from circumstellar dust
shells is determined by the composition of the dust mixture and the properties
of the different dust species. Here we concentrate on the special question 
whether the unusual feature at 10\,$\mu$m detected in a number of S-stars
could result from solid SiO. Since this feature is the most prominent dust
feature in the objects where it has been detected so far \citep{Hon09}, it 
appears also to be the most abundant dust species in these objects, apart from
possible contributions to opacity from  dust species with structure-less pure
continuous extinction like, e.g., iron dust grains. Therefore we concentrate
on models with solid silicon monoxide as the sole dust component. 

In our model calculations we assume the opacity in the shell to be completely
determined by the dust components; no contribution of the gas phase is 
considered. The dust particles are assumed to be small spheres of radius $a$.
The absorption and scattering coefficients of the different dust components, 
characterized by an index $i$, then are given by
\begin{equation}
\varkappa_i^{\rm abs, sc}=\varrho\,
{3\over4}\,{A_i\epsilon_i\over(1+4\epsilon_{\rm He})\rho_{{\rm d},i}}
\ {\cal Q}_{i}^{\rm abs, sc}\,f_i\,.
\label{AbsScCoeff}
\end{equation}
The quantity $A_i$ is the molecular weight corresponding to the chemical 
formula of the condensed phase, $\epsilon_i$ is the elemental abundance of a
characteristic key element that is required to form the condensed phase, and
$f_i$ is the fraction of the atoms of this element that is really condensed
into the solid phase, $\rho_{{\rm d},i}$ is the mass-density of the solid. The
quantities ${\cal Q}_{i}^{\rm abs, sc}$ are the absorption and scattering
efficiencies, divided by the particle radius~$a$. These quantities are 
calculated by means of Mie-theory \citep[see][]{Boh83} from the dielectric
function of the dust material. The basic data used in the calculation of
dust extinction are given in Table~\ref{TabDustSpec}.

With respect to the particle shape we consider only spherical grains. If the
grains would be non-spherical, the absorption band profiles of dust grains 
would be distorted and their centre would be shifted compared to the case of
spherical grains. However, SiO grains formed in laboratory condensation 
experiments from  the gas phase generally seem to be almost spherical 
\citep[][ their Fig. 1a]{Kam04}. Since there is no obvious reason why this
should be different for the tiny grains condensed in a circumstellar shell and
for nanometre sized laboratory  condensed grains, this leaves no room for 
introducing a distribution of grain shapes like, e.g., the popular continuous
distribution of ellipsoids \citep[see, e.g.,][]{Boh83}, in order to account 
for possible non-sphericity effects.

\begin{figure}

\includegraphics[width=\hsize]{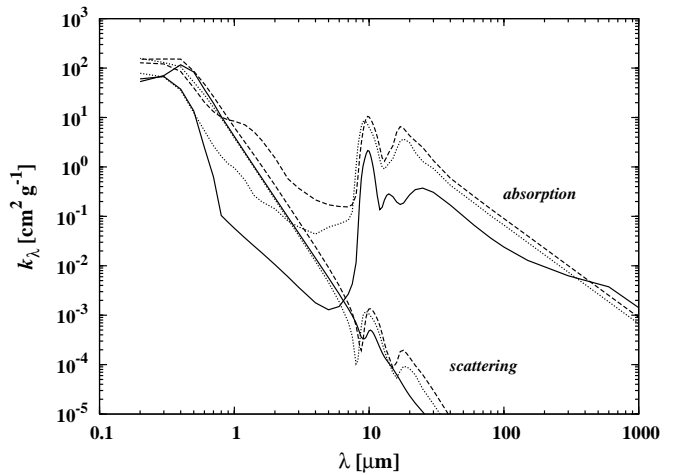}

\caption{Absorption and scattering coefficient per unit mass of some dust species
in a dusty gas with standard cosmic element abundances. Full line: solid SiO. 
Dashed line: amorphous olivine: Dotted line: amorphous pyroxene.}

\label{FigAbsCoeff} 
\end{figure}

With respect to the particle size we assume the canonical value of $a=0.1\,\mu$m
that seems to be typical for dust grains in circumstellar shells
\citep[e.g.][]{Kru68,Ser01}. Without calculation of a complete model for
the dust shell including hydrodynamics and dust growth the grain radius cannot
be specified more precisely. Fortunately, for the far infrared spectral region
the opacity does not depend on the grain size $a$ since circumstellar grains
are in any case much smaller than the wavelength $\lambda$ in the far infrared.
Only for the absorption of stellar radiation there may be a slight dependency
on grain size since in the optical and near UV spectral region the grains may
satisfy only marginally the condition $2\pi a/\lambda\ll1$ that particles can
be considered as small. 

For the purpose of model calculations of dust shells of cool giant stars one
needs optical data for the dust species at least in the wavelength range from
about 0.4\,$\mu$m to about 100\,$\mu$m in order to cover (i) the wavelength
range of the stellar radiation field from the ultraviolet to the mid infrared,
and (ii) to cover the wavelength range of dust emission from the shell that
ranges from the near infrared to the sub-mm region. Our own measurements
discussed in Sect.~\ref{SectOptPropSiO} cover only the mid to far infrared
wavelength range. Therefore we have to augment our data by data from
other sources to cover also the spectral region from near infrared to UV.

For the visible to far-ultraviolet spectral regions ($\lambda\le0.8\,\mu$m) we
use data for solid SiO listed in \citet{Pal85}. The real part of the complex
index of refraction of our infrared extinction data and the data given in
Palik nicely fit in the near infrared where they overlap. The imaginary part
becomes very small in the wavelength region below $8\,\mu$m in our results and 
it is omitted in the range between 8 and $0.8\,\mu$m in the listing of Palik. In
this region the absorption coefficient of solid SiO would become many orders of
magnitude smaller than the scattering coefficient (or would vanish at all if
data of Palik are used). In order to avoid numerical problems in solving the
radiative transfer problem we arbitrarily increase the absorption in the
wavelength range 0.8 to 8\,$\mu$m by assuming that the dust material contains
tiny inclusions of pure iron particulates with a volume filling factor of
$f_{\rm V}=10^{-3}$ and calculating the average dielectric function 
$\langle\varepsilon\rangle$ for this composite material from the Maxwell-Garnett
mixing rule \citep[cf., e.g.,][]{Boh83}
\begin{equation}
\langle\epsilon\rangle={(1-f_{\rm V})\epsilon_{\rm m}+f\beta\epsilon_{\rm inc}
\over1-f_{\rm V}+f_{\rm V}\beta}
\end{equation}
where
\begin{equation}
\beta={3\epsilon_{\rm m}\over\epsilon_{\rm inc}+2\epsilon_{\rm m}}\,.
\label{DefBetaMG}
\end{equation}
Here $\epsilon_{\rm m}$ is the bulk dielectric function of the matrix material
(solid SiO in our case) and $\epsilon_{\rm inc}$ the bulk dielectric function
of the material forming the inclusions (solid iron in our case). 

This kind of additional absorption by impurities has no influence on the 
calculated temperature of dust grains since this is determined by the strong
absorption at $\lambda<0.8\,\mu$m and also has no influence on the calculated
radiation field. It is unlikely that any real dust material is pure and ideally
transparent, and nano-sized iron inclusions are a fundamental constituents of,
e.g., GEMS \cite[e.g.][]{Bra03}. 

The resulting mass-absorption and scattering coefficients of small particles of
solid SiO are shown in Fig.~\ref{FigAbsCoeff}. They are calculated according to
Eq.~(\ref{AbsScCoeff}) using the data given in Table~\ref{TabDustSpec}, and 
calculating optical constants for SiO according to the Brendel-model for the
dielectric function of amorphous materials, Eq.~(\ref{ModlLorentzAm}), using 
constants from Table~\ref{ConstEpsSiO}, and following the procedure described 
above.  

\citet{Brendel92} have shown that the integral in Eq.~(\ref{ModlLorentzAm})
can be solved in terms of the complex probability functions and Gaussians. We
did not make use of this in our opacity calculations but preferred to evaluate
the integrals numerically since evaluation of the analytic expressions seems not
to offer any particular advantage due to their complicated nature.

For comparison also mass-absorption and scattering coefficients of amorphous
olivine and pyroxene are shown,%
\footnote{%
Data for silicates with $x=0.7$, from the Jena-St. Petersburg data basis,
accessible via: 
{\tt\tiny http:/www.mpia-hd.mpg.de/HJPDOC/}
}
that are the main dust species in circumstellar dust of M-Stars. It is seen that
solid SiO indeed shows a prominent absorption feature centred around about 
10\,$\mu$m and only some weak  structure on the long wavelength side.

\begin{table}

\caption{Basic parameters of the dust-shell models}

\begin{tabular}{l@{\hspace{.9cm}}l@{\hspace{.9cm}}r@{\hspace{.5cm}}l}
\hline\hline
\noalign{\smallskip}
Star   & $T_{\rm eff}$      & 2\,700              & K                   \\
       & $L_*$              & $1\times10^4$       & L$_{\sun}$           \\
       & $R_*$              & $3.18\times10^{13}$ & cm                  \\
       &                    & = 534               & R$_{\sun}$           \\
\noalign{\smallskip}
Wind   & $v_{\rm exp}$      & 10                  & km\,s$^{-1}$        \\
       & $\dot M$           & $1\times10^{-6}$ --- $3\times10^{-5}$
       & M$_{\sun}$\,a$^{-1}$ \\
\noalign{\smallskip}
Dust   & $\kappa^{\mbox{\tiny abs}}$ & amorphous SiO  &    \\
       & $f_{\rm SiO}$      & 0.5                 &    \\
\noalign{\smallskip}
Shell  & $R_{\rm i}$        & depends on $\dot M$ &               \\
       & $R_{\rm a}$        & $1\times10^5$       & R$_*$               \\
       & $T_{\rm c}$        & 750                 & K                   \\
\noalign{\smallskip}
\hline
\end{tabular}

\label{TabShellParm}
\end{table}

\subsection{Model calculations}

Using the above opacity description we have calculated radiative transfer
models of circumstellar dust shells with solid SiO as dust component for a
range of mass-loss rates in order to determine the spectral features that
would result from solid silicon monoxide. The essentials of the model of the
dust shell and of the radiative transfer calculations are briefly described in
Appendix~\ref{SectCalcRadTraMod}.

A fundamental parameter of the model is the location of the inner boundary of the
dust shell, $R_{\rm i}$, that depends on the mechanism by which the dust forms
from the gas-phase by nucleation and subsequent grain growth. About the
details of these processes almost nothing is known and we take recourse to the
simple assumption that the dust suddenly appears at some prescribed dust 
temperature $T_{\rm c}$ and the fraction $f_{\rm SiO}$ of material condensed
into the solid phase is constant across the dust shell. This is the kind of 
approximation on which most existing models of circumstellar dust shells are
based on. For the temperature $T_{\rm c}$ we assume for silicon monoxide dust a
value of  750\,K because the temperature where solid SiO becomes stable against
vaporisation under conditions in circumstellar dust shells is about 800\,K, 
see Sect.~\ref{SectSstars}, and because in dust shells this temperature
probably is somewhat lower if some super-cooling of the outflowing material
should be required for the onset of nucleation and condensation. The old
results of \citet{Gai86} seem to indicate a temperature for onset of SiO
condensation as low as 600\,K, but this cannot be maintained since that
calculation is based on older vapour pressure measurements of solid SiO for
which it is now known that they have substantially to be revised downwards
\citep{Fer08,Kle09}, which increases the condensation temperature.
  
\begin{figure}

\includegraphics[width=.9\hsize]{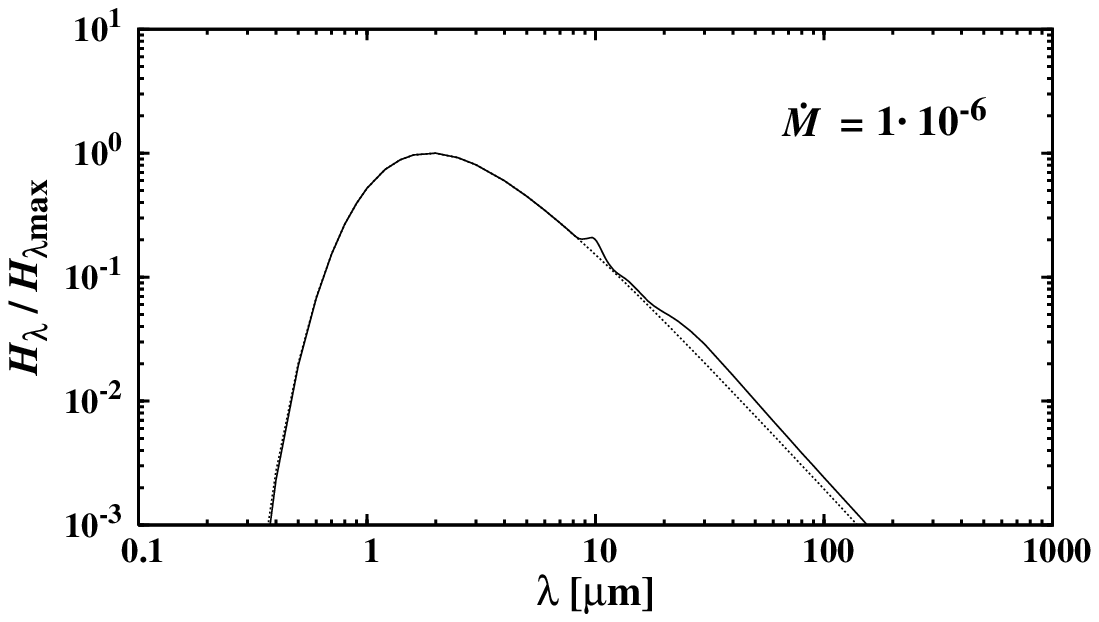}

\includegraphics[width=.9\hsize]{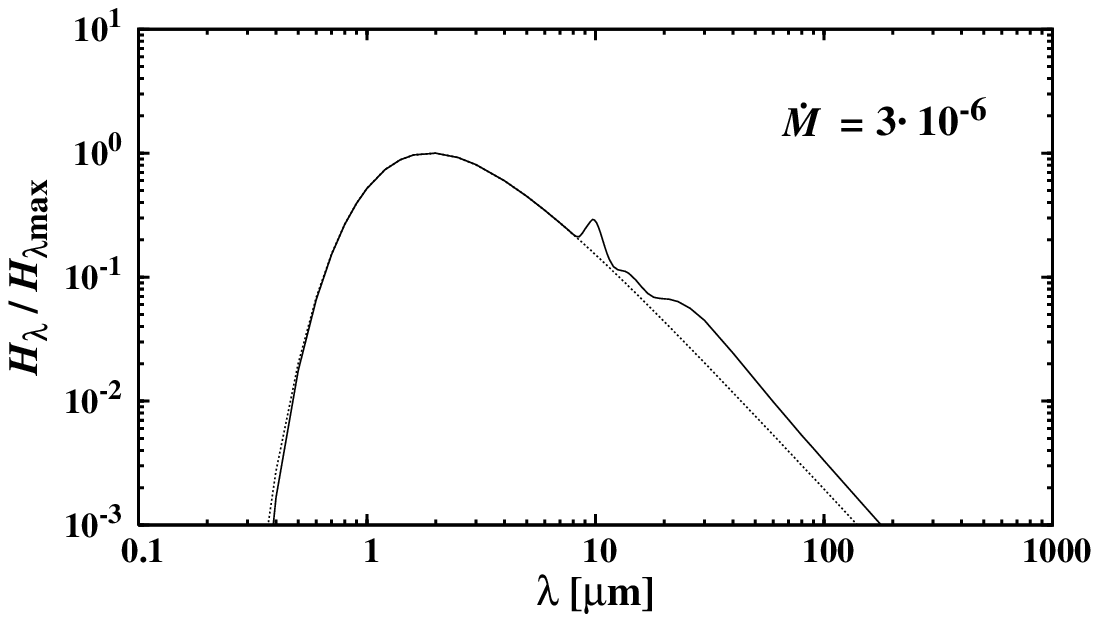}

\includegraphics[width=.9\hsize]{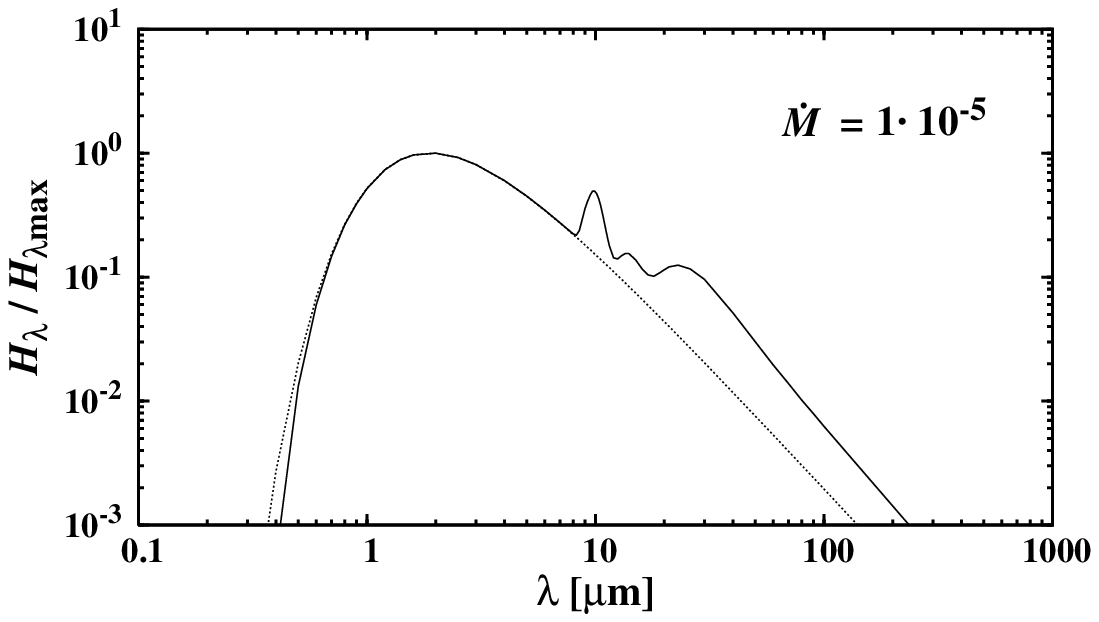}

\includegraphics[width=.9\hsize]{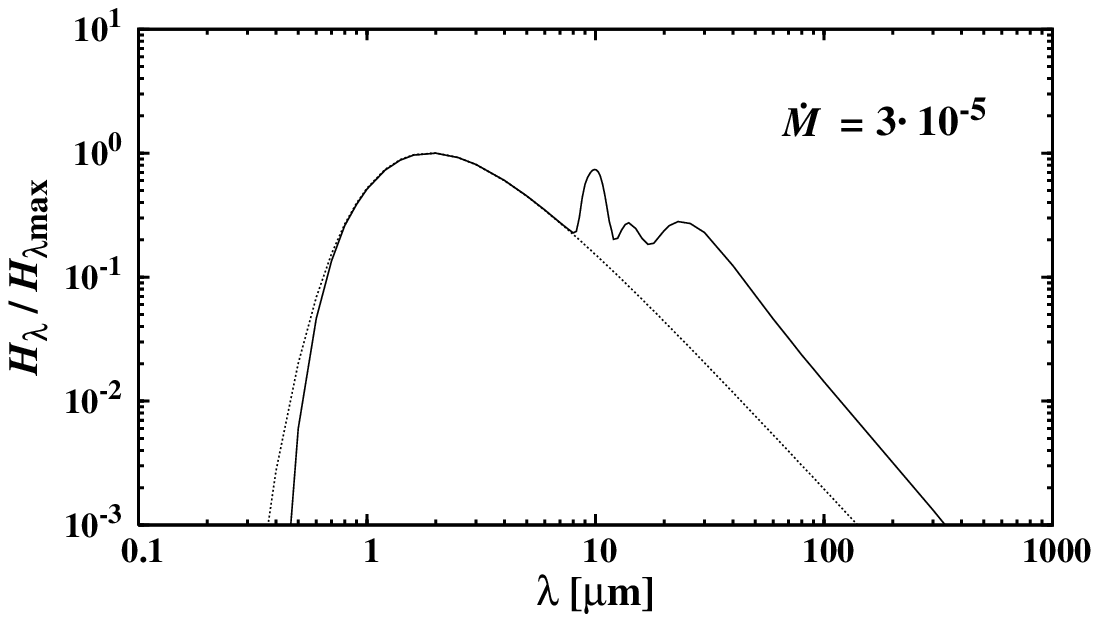}

\caption{Synthetic spectra of models of circumstellar dust shells with solid
silicon monoxide dust particles for different mass-loss rates from
$\dot M=1\times10^{-6}\,\rm M_{\sun}\,yr^{-1}$ to
$\dot M=3\times10^{-5}\,\rm M_{\sun}\,yr^{-1}$. Dotted line: stellar radiaton. 
}

\label{FigSpect} 
\end{figure}

For the basic parameters $T_{\mathrm{eff}}$, L$_*$, $V_{\mathrm{exp}}$ we use
representative average values, since we do not intend to model in\-di\-vidual 
stars but to perform only some explorative calculations. For the effective
temperature of the central star we chose a value of 2\,700\,K. This seems to
be a representative value for S stars \citep[e.g.][]{Ker96,Ker99}.
The luminosity is chosen to be $L=10^4\,{\rm L}_{\sun}$, which is typical for
stars at the uppermost part of the TP-AGB, where one expects AGB-stars to
pass through the stage of S-stars. Observed values given in the literature for
stars with substantial mass-loss rate are of this order of magnitude or
slightly less \citep[e.g.][]{Jor98,Gro98,Ram06}. The observed expansion 
velocities $V_{\mathrm{exp}}$ of S-stars vary between a few km\,s$^{-1}$ and
about 20\,km\,s$^{-1}$ and are typically between 5\,km\,s$^{-1}$ and 
10\,km\,s$^{-1}$ \citep{Gro98,Ram06}. We use a value of 10\,km\,s$^{-1}$.
The observed mass-loss rates of S-stars vary over a large range 
between values as low as a few times $10^{-8}\,\rm M_{\sun}\,yr^{-1}$ to
several times $10^{-5}\,\rm M_{\sun}\,yr^{-1}$ \citep[e.g.][]{Gro98,Ram06}. We 
consider $\dot M$ as free parameter and calculate models for different values
of the mass-loss rate.

The basic model parameters of the calculations are shown in 
Table~\ref{TabShellParm}.

\begin{figure}

\includegraphics[width=\hsize]{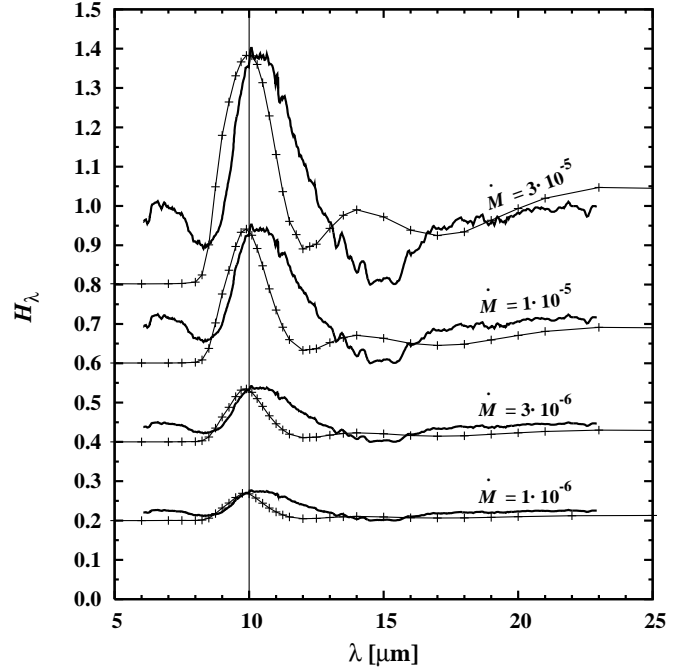}

\caption{Emitted energy flux from star and dust shell, subtracted by 
black-body emission from the star (lines with crosses) for different mass-loss
rates (in units M$_{\sun}$\,yr$^{-1}$), and averaged emission profile of
S-stars from \citet{Hon09}. The S-star line profile is scaled such that the
maxima of observed and calculated profiles coincide; for clarity, the spectra
are shifted by arbitrary amounts.
}

\label{FigSiOFeature}
\end{figure}

\subsection{Results for spectral energy distribution}

Models are calculated for mass-loss rates ranging  from $\dot M=
1\times10^{-6}\,\rm M_{\sun}\,yr^{-1}$ to $\dot M=3\times10^{-5}\,\rm M_{\sun}
\,yr^{-1}$ with solid silicon monoxide as the sole opacity source. Figure 
\ref{FigSpect} shows the resulting spectral energy distribution of the
radiation emitted by the star and its dust shell. Since we assumed for the
star a black-body radiation field, the strong structure of cool stellar 
spectra resulting from molecular absorption bands is missing in
our models. This  simplifies the identification of the dust features in 
the spectrum. Emission from warm dust is seen only at wavelengths longer than 
8\,$\mu$m since the absorption coefficient of solid SiO is small at shorter
wavelengths, see Fig.~\ref{FigAbsCoeff}. The prominent absorption band around 
10\,$\mu$m is clearly seen as a strong emission feature in all of the spectra.
Some structure is seen in the spectrum at longer wavelengths but there is no
strong emission feature around 18\,$\mu$m. 

This is similar to what is found in a number of S-star spectra. Figure 
\ref{FigSiOFeature} shows a more close-up view of our synthetic spectra in
the wavelength region between 5\,$\mu$m and 25\,$\mu$m. For clarity the black
body radiation field of the star is subtracted from the synthetic spectra and
the resulting emission profiles are shifted relative to each other. These
emission profiles reflect the peculiar characteristics of the emission
coefficient of amorphous solid silicon monoxide: A strong emission band 
centred around 10\,$\mu$m and only weak to negligible structure in the 
15\,$\mu$m to 25\,$\mu$m wavelength region. Figure~\ref{FigSiOFeature} also
shows one of the two different profiles of the emission bands around
10\,$\mu$m observed in S-stars where no or only a very weak 18\,$\mu$m feature
is observed. The data are  taken from Fig.~6 of \citet{Hon09}. The profiles in
Fig.~\ref{FigSiOFeature} corresponds to the lower one in Fig.~6 of 
\citet{Hon09}. This profile is scaled such that the peak value coincides with
that of the synthetic emission profiles in our models. 

The upper profile of Fig.~6 of \citet{Hon09} is not considered since it peaks
at a significantly higher wavelength. It may result from a different kind of
material or a complex superposition of emission bands from a number of
materials. 

\subsection{Discussion}

The calculated line profiles show some similarity with the average profile
derived in \citet{Hon09} which peaks at 10\,$\mu$m. The peak position does not
match exactly but probably within the limit of errors
with which such averaged profiles can be determined. The broad extension to
the long wavelength side of the observed profile, however, does not agree with 
the shape of our profiles. The strong asymmetry of the observed profile is
conspicuous. This may have a number of reasons. Either the observed profile is
the superposition of the strong 10\,$\mu$m emission feature of solid silicon
monoxide with one or more absorption bands of additional dust materials centred
around somewhat longer wavelengths, or the properties of solid silicon monoxide
deposited under conditions in circumstellar shells differ from that of the
material obtained in our laboratory experiments by CVD.

With respect to a contribution by some other materials the most likely 
candidates are hibonite (CaAl$_6$O$_{19}$) with a strong absorption band
centred on $12.3\,\mu$m, spinel (MgAl$_2$O$_4$) with a strong absorption band
centred on 13\,$\mu$m, or corundum Al$_2$O$_3$ with a strong and broad band
peaking at 11\,$\mu$m. Such materials probably would start to condense at
higher temperature than solid SiO because of their lower vapour pressure. They
would use up only small amounts of the oxygen because of the much lower element
abundances of Al and Ca compared to that of Si. This would not noticeably
compete with SiO condensation under conditions where oxygen is scarce. 
Explorative calculations showed, however, that the hibonite \citep[optical data
from][]{Mut02} and spinel \citep[optical data from][]{Pal85} band would always
be seen as a separate band, which seems not to be found in spectra of S-stars 
with a strong 10\,$\mu$m feature. The rather broad corundum feature 
\citep[optical data from][]{Koi95} also  cannot explain the strong asymmetry 
of the observed profile, but it would fill up the dip of the emission profile
due to solid SiO at about 12.5\,$\mu$m.

Also forsterite may contribute somewhat to the observed profile and increase
its width and asymmetry, because if the oxygen abundance is not close to the
critical abundance limit $\epsilon_{\mathrm{O,crit}}=\epsilon_{\mathrm{C}}-
\epsilon_{\mathrm{Si}}+\epsilon_{\mathrm{S}}$ where O is completely bound by CO
and by that fraction of Si that is bound in SiO and not in SiS, then some
oxygen is left over to form silicates. However, it seems not to be possible to 
reproduce the observation with a mixture of solid SiO and silicates, because if
more than a few per cent of the Si forms a silicate, the 18\,$\mu$m feature
becomes clearly visible. On the other hand, additional formation of silicates
besides solid SiO could explain the sources with a marked 10\,$\mu$m feature
but an only weak 18\,$\mu$m feature, which are frequently found within the
population of S-stars. 

Possibly one could reproduce the observed band structure of the stars under
consideration by a mixture of the oxides and silicates with solid SiO, but this
is beyond the scope of this paper. 

Another possibility for the strong asymmetry of the observed profile may be
that the material formed in stellar outflows is more strongly disordered than
the material formed in our laboratory experiments. A significantly stronger
damping constant $\gamma_j$ would also result in an asymmetric and  much
broader line profile. However, test calculations with strongly increased values
of $\gamma_j$ showed that the observed band shape cannot be explained in this
way.
  
Finally some remarks are in place on the structure of SiO. At low temperatures
this material slowly decomposes into silicon nanoclusters embedded in a SiO$_2$
matrix \citep{Kam04}. The material called silicon monoxide then in truth is an
intimate mixture of two different phases which is inhomogeneous on scales of a
few nanometres. The absorption peak then shifts to the characteristic peak of 
SiO$_2$ at 9.2\,$\mu$m. Clearly, such a decomposition did not occur for our
silicon monoxide films and would also be of minor importance in circumstellar
shells if solid SiO would be the carrier of the 10\,$\mu$m feature.
Decomposition times have been measured for the closely related material
Si$_2$O$_3$ in the laboratory by, e.g., \citet{Nut84}. If this can be taken as
representative for solid SiO the decomposition would be sufficiently slow
at the temperatures of interest ($\lesssim750$¸\,K) that solid SiO can survive
for a sufficiently long time to be observable.


\section{%
Conclusions}

This paper reports on laboratory studies of the infrared optical properties of
solid silicon monoxide. The dielectric function $\varepsilon$
is derived from transmission measurements of thin films obtained by evaporation
of commercially available solid SiO and depositing its vapour on a cold 
substrate. A Brendel oscillator model is fitted to the results of the 
transmission measurements to determine $\varepsilon(\omega)$.

Solid silicon monoxide is a material that has been speculated at different 
occasions to be involved in the formation process of silicate dust in
circumstellar environments like accretion disks around protostars or dust shells
around AGB stars. In this paper we study the possibility that solid silicon 
monoxide forms in circumstellar dust shells a separate dust component if 
insufficient oxygen is available to build the SiO$_4$-tetrahedrons of normal
silicate dust material. This would apply for instance to S-stars with a C/O 
abundance ratio very close to unity. 

The characteristic property of such SiO dust would be a broad and structure-less
emission feature centred on about 10\,$\mu$m resulting from Si-O bond stretching
vibrations, similar to the 9.7\,$\mu$m feature of amorphous silicates, but a
missing 18\,$\mu$ feature resulting from O-Si-O bending modes in a 
SiO$_4$-tetrahedron. Such peculiar cases have, indeed, been detected for some
S-stars \citep{Hon09}. We propose that we see in these stars solid silicon
monoxide as an abundant dust component. Radiative transfer calculations for
circumstellar dust shells have been performed using our new data on the
dielectric function of solid SiO to calculate the emission band structure
due to SiO dust.

The resulting emission feature at 10\,$\mu$m is compared with one of the
two average emission profiles derived in \citet{Hon09}. The profile of the
emission feature obtained in the model calculation peaks at about just the 
same wavelength as the observed feature, which makes its identification as due
to solid SiO very likely. However, the observed profile is much more extended
to longer wavelengths than the synthetic profile which either may result from
blending the band from solid SiO with emission bands from a number of
additional minor dust species (corundum, hibonite, ...) or shows that the
laboratory produced amorphous SiO films have a somewhat different lattice
structure than the material condensed in stellar outflows. 

More work is required to explain the asymmetry of the observed peculiar 
10\,$\mu$m feature in some S-stars before the identification of solid silicon
monoxide as carrier material of this feature can be considered as safe; 
but in any case this explanation can presently be considered as the most likely
one.


\begin{acknowledgements}
This work was supported in part by `Forschergruppe 759' and special research progamm
SPP 1385  which both are supported by the `Deutsche Forschungs\-gemeinschaft (DFG)'. 
\end{acknowledgements}


\begin{appendix}

\section{%
Radiative transfer model}
\label{SectCalcRadTraMod}

Model spectra of S-stars are calculated using a simple code for modelling
radiative transfer in circumstellar dust shells. The basic assumptions on the
structure of the dust shell follow the conventional assumptions: A spherically
symmetric mass distribution around the central star, with inner radius 
$R_{\rm i}$ and outer radius $R_{\rm a}$. The outflow is assumed to be stationary
and the outflow velocity $v_{\rm exp}$ to be independent of distance $r$ from the
centre. The radial distribution of mass density in this case is 
\begin{equation}
\varrho(r)={\dot M\over4\pi r^2v_{\rm exp}}\,.
\end{equation}
Here $\dot M$ is the mass-loss rate that is also assumed to be constant.

With given opacities (see Sect.~\ref{SectOpac}), the radiative transfer equation
in spherical symmetry 
\begin{equation}
\pd{I_\nu}r+{1-\mu^2\over r}\pd{I_\nu}{\mu}=-\varkappa^{\rm ext}
\left(I_\nu-S_\nu\right)
\end{equation}
is solved by the so called $p$-$z$-method \citep[cf.][]{Mih78}. The source
function  $S_\nu$ and the total extinction coefficient $\varkappa^{\rm ext}$ are
\begin{eqnarray}
S_\nu\!&=&\!\!\sum\limits_{i=1}^I\left\{{\varkappa_i^{\rm abs}\over \varkappa^{\rm ext}}
B_{i,\nu}(T_i)\!+\!{\varkappa_i^{\rm sc}\over \varkappa^{\rm ext}}
{3\over 8}\left[(3-\mu^2)J_\nu\!+\!(3\mu^2-1)K_\nu\right]\right\}\\
\varkappa^{\rm ext}\!\!&=&\!\!\sum\limits_{i=1}^I\left(\varkappa_i^{\rm abs}+
\varkappa_i^{\rm sc}\right)\,.
\end{eqnarray}
The sum runs over all dust species.  The angular distribution of the scattering
term is that for small particles. The temperatures $T_i$ of the dust species
are determined such as to satisfy radiative equilibrium 
\begin{equation}
\int_0^\infty{\rm d}\nu\,\varkappa_i^{\rm abs}\left[J_\nu-B_\nu(T_i)\right]=0
\end{equation}
for each species by applying a
Uns\"old-Lucy temperature correction procedure \citep[see][]{Luc64} adapted to 
the spherically symmetric case. This requires an iteration procedure for the 
determination of all temperatures $T_i$. This iteration is combined with a simple
iteration scheme (successive over-relaxation) with respect to $J_\nu$ and 
$K_\nu$ in the scattering contribution of the source function.

The inner radius $R_{\rm i}$ of the dust shell is fixed by the requirement,
that the most stable of the dust species appears at some prescribed temperature
$T_{\rm c}$. Since it is not known in advance at which radius this condition
is satisfied, an additional iteration procedure is required to determine 
$R_{\rm i}$. The outer radius $R_{\rm a}$ is always taken at $10^5$
stellar radii.

The radiation field of the central star is approximated by a black body
radiation field.


\section{Relation to Voigt function}

\label{SectVoigt}

We show here how the Brendel oscillator model is related to the more familiar
Voigt function for the profile of damped spectral lines.

A single oscillator of the Brendel model contributes to the dielectric function
a term
\begin{equation}
X_j(\omega)={1\over2\pi\sigma_j}\int\limits_0^\infty{\rm d}z\,
{\rm e}^{-(z-\omega_{0,j})^2/2\sigma_j^2}
\,{\omega_{{\rm p},j}^2\over z^2-\omega^2-{\rm i}\gamma_j\omega}
\label{OszBrendel}
\end{equation}
or, after splitting this into its real and imaginary part, the terms
\begin{eqnarray}
X_{j,\rm r}&=&{1\over2\pi\sigma_j}\int\limits_0^\infty{\rm d}z\,
{\rm e}^{-(z-\omega_{0,j})^2/2\sigma_j^2}
\,{\omega_{{\rm p},j}^2(z^2-\omega^2)\over 
(z^2-\omega^2)^2-\gamma_j^2\omega^2}
\\
X_{j,\rm i}&=&{1\over2\pi\sigma_j}\int\limits_0^\infty{\rm d}z\,
{\rm e}^{-(z-\omega_{0,j})^2/2\sigma_j^2}
\,{\omega_{{\rm p},j}^2\gamma_j\omega\over (z^2-\omega^2)^2-\gamma_j^2\omega^2}
\,.
\end{eqnarray}
If $\gamma\ll\omega$, as in our case, the Lorentz profile as function of $z$ is
different from zero  in an only small interval around $\omega$. Letting
$\Delta z=z-\omega$, $\vert\Delta z\vert\ll\omega$ one obtains by retaining 
only terms of lowest order in $\Delta z$
\begin{eqnarray}
X_{j,\rm r}&=&{1\over2\pi\sigma_j}\int\limits_0^\infty{\rm d}z\,
{\rm e}^{-(z-\omega_{0,j})^2/2\sigma_j^2}
\,{\omega_{{\rm p},j}^2\over2\omega}{(z-\omega)\over (z-\omega)^2-\frac14\gamma_j^2}
\\
X_{j,\rm i}&=&{1\over2\pi\sigma_j}\int\limits_0^\infty{\rm d}z\,
{\rm e}^{-(z-\omega_{0,j})^2/2\sigma_j^2}
\,{\omega_{{\rm p},j}^2\over4}{\gamma_j\over (z-\omega)^2-\frac14\gamma_j^2}\,.
\end{eqnarray}
If we compare the imaginary part with definition 
\begin{equation}
H(\alpha,v)={\alpha\over\pi}\int_{-\infty}^{+\infty}{\rm d}x\,{\rm e}^{-x^2}
{1\over(x-v)^2+\alpha^2}
\end{equation}
of the Voigt function \citep[see][for details]{Mih78} we recognize that the
imaginary part of one oscillator in the Brendel model corresponds, except for
a constant factor, to the Voigt function with parameters
\begin{eqnarray}
v&=&{\omega-\omega_0\over\sqrt2\sigma}\\
\alpha&=&{\gamma^2\over8\sigma}\,.
\end{eqnarray} 
Since the Voigt function is the real part of the complex error function
\citep[e.g.][who also gives a FORTRAN routine for calculating the complex
probability function]{Hum82}, the Brendel oscillators are also directly related to
the complex probability function.

\end{appendix}


\end{document}